\documentclass[aps,pre,twocolumn,floatfix]{revtex4-1}
\usepackage{graphicx}
\usepackage{float}
\usepackage{commath,amssymb,amsmath,stmaryrd,tabularx}

\DeclareMathOperator{\csch}{csch}
\begin{document}
\title{Pulsatile Driving Stabilizes Loops in Elastic Flow Networks}
\author{Purba Chatterjee, Sean Fancher, and Eleni Katifori}
\affiliation{Department of Physics and Astronomy, University of Pennsylvania, Philadelphia, Pennsylvania $19104$, USA}

\begin{abstract}

Existing models of adaptation in biological flow networks consider their constituent vessels (e.g. veins and arteries) to be rigid, thus predicting a non physiological response when the drive (e.g the heart) is dynamic. Here we show that incorporating pulsatile driving and properties such as fluid inertia and vessel compliance into a general adaptation framework fundamentally changes the expected structure at steady state of a minimal one-loop network. In particular, pulsatility is observed to give rise to resonances which can stabilize loops for a much broader class of metabolic cost functions than predicted by existing theories. Our work points to the need for a more realistic treatment of adaptation in biological flow networks, especially those driven by a pulsatile source, and provides insights into pathologies that emerge when such pulsatility is disrupted in human beings.
\end{abstract}
\maketitle
\section{Introduction}
The structure of physiological transport networks such as animal vasculature and leaf venation, has important consequences for biological functionality, and as such has elicited considerable scientific interest over the years. In particular, looped network architectures, ubiquitous in biology, are beneficial for mitigating vessel damage and optimizing responses to source fluctuations \cite{Katifori2010,Corson2010,Kaiser2020}. Network remodeling \cite{Pries1998,Pries2008}, also known as adaptation, is now understood to proceed by optimizing the total energy dissipation in the network, subject to some metabolic cost \cite{Hu2013,Chang2019,Gounaris2021,Kramer2021}. Such metabolic costs can generally be described by a power law ($K^\sigma$), where $K$ is vessel conductivity and $\sigma$ is a system specific parameter. Existing theories of adaptation in flow networks predict a critical transition at $\sigma=1$, with a structure with many loops for $\sigma<1$ and one which is a loop-less tree for $\sigma>1$ \cite{Banavar2000,Durand2007,Bohn2007,Katifori2010,Corson2010}. However the scope and generality of this prediction remains to be investigated in the light of biologically relevant dynamical considerations.

Previous studies on adaptation consider vessels to be rigid, leading to the assumption that modulations in flow boundary conditions are instantaneously propagated to individual network elements at all times. However, vessel compliance and fluid inertia have been shown to generate a finite timescale of information transfer from the sources to the bulk \cite{Fancher2022, Fancher2021}, leading to trade-offs between energy efficiency on one hand and mechanical response to sudden changes in the steady state dynamics on the other \cite{Fancher2022}. Moreover, while fluctuating sources and sinks have been implemented within the framework of network adaptation \cite{Hu2013,Katifori2010,Hu2012,Grawer2014}, the effect of deterministic pulsatile driving at the source is largely unexplored . This is an important consideration for biological transport networks, many of which rely on pulsatility to maintain fluid pressure. The most prominent example of this is mammalian vasculature, with the periodic beating of the heart muscle introducing pulsatile components into blood flow.

In this Letter, we investigate the effect of both pulsatile driving and the internal spatio-temporal dynamics of elastic vessels on the adaptation of a simple one-loop flow network. Depending on the lengths of the vessels in relation to each other and to a characteristic length scale over which pulsatility is damped, resonant frequencies are shown to exist, which amplify energy dissipation and stabilize loops for a much broader class of metabolic cost functions than predicted by existing theories of adaptation. Our results emphasize the need for a more sophisticated treatment of adaptation in order to correctly predict the steady state structure of more complicated biological transport networks and might be key in explaining the development of vascular malformations in patients with artificial hearts \cite{hexnerPNAS2020,Bauerle2020}.
\begin{figure}
\begin{center}
\includegraphics[scale=0.39]{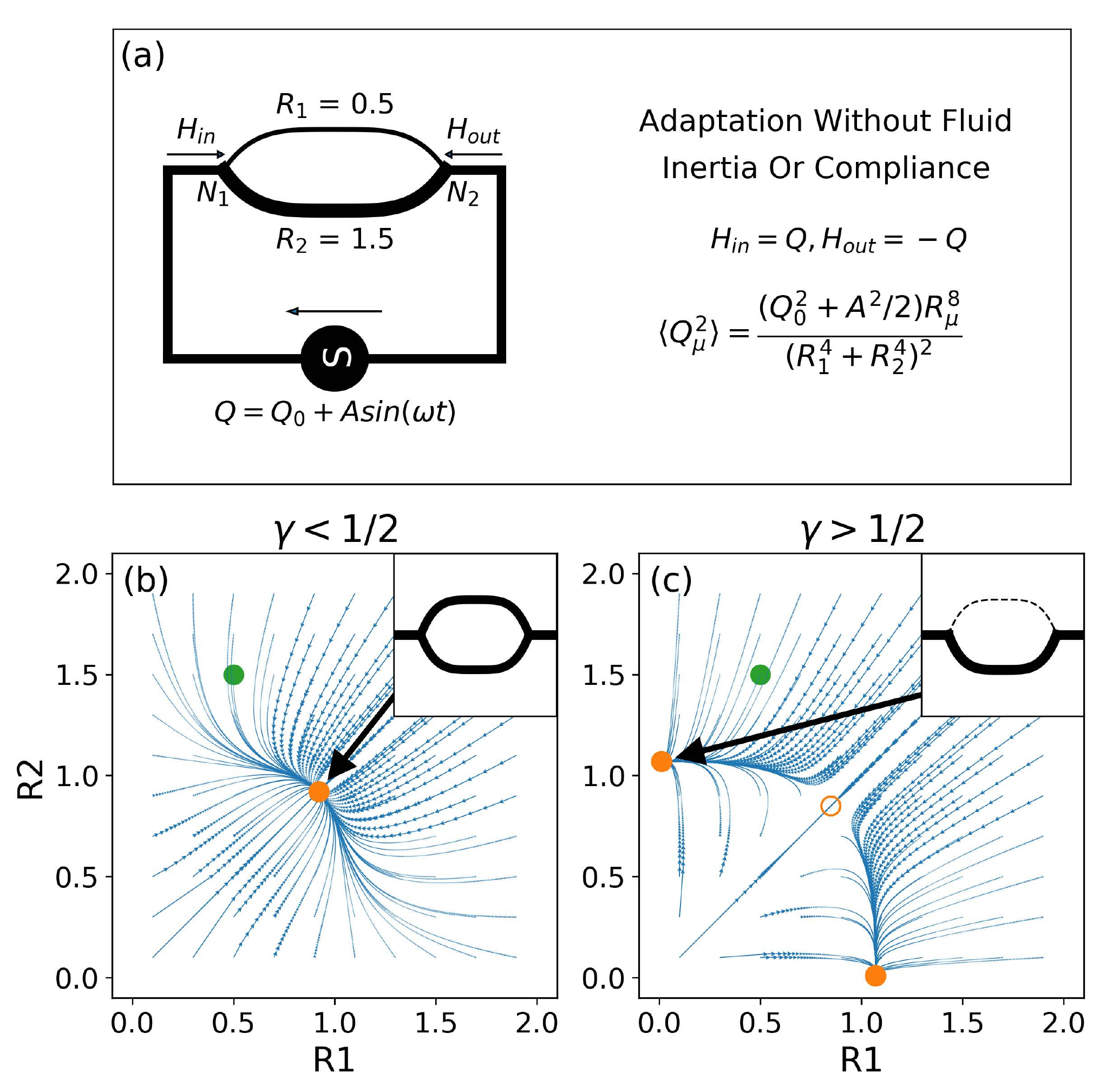}
\end{center}
\caption {\small{(color online) (a) Minimal model with $L_1=L_2=L$, and the vessel mean-squared current in the absence of fluid inertia and vessel compliance. (b,c) Flow diagrams for AE in $R_1 - R_2$ space, $a=b=L=1$. Insets show radii at steady state for the chosen initial condition (green solid circle) depicted in (a). Sinks at steady state denoted by solid orange circles, and saddle points by hollow orange circles.}}
    \label{fig1}
\end{figure}

\section{The Adaptation Equation}
Our minimal network consists of two vessels connecting the nodes $N_1$ and $N_2$ (Fig.~\ref{fig1}(a)), subject to pulsatile driving at the frequency $\omega$. The vessel radii $R_{\mu}$, $\mu\in\lbrack 1,2\rbrack$, change in response to the local current $Q_{\mu}$ over time $t'$, according to the Adaptation Equation (AE)
\begin{equation}\label{eq1}
\frac{dR_{\mu}}{dt'}=\frac{a\langle Q_{\mu}^2\rangle^\gamma}{R_{\mu}^3}-bR_{\mu},
\end{equation}
where $a$ and $b$ are constants, and $\gamma$ is a parameter associated with the metabolic cost. 
Assuming Poiseuille flow ($K_{\mu} \propto R_{\mu}^4$), this AE is identical to a general adaptation rule that has been used to model the dynamics of hydraulic vessel conductivities ($K_{\mu}$) during animal vascular development and the slime mold Physarum polycephalum \cite{Bauerle2020,Hacking1996,RollandLagan2005,Hu2013,Berkel2013,Ronellenfitsch2016,Ronellenfitsch2019}. Each vessel adapts through a local positive feedback, expanding in radius when the current through it is large, and shrinking at the characteristic timescale $b^{-1}$ when it is small. The steady states of the AE correspond to the critical points of the optimization functional
\begin{equation}\label{eq2}
E=\sum_{\mu}L_{\mu}\frac{Q_{\mu}^2}{K_{\mu}} +\beta\Bigg(\sum_{\mu} L_{\mu} K_{\mu}^\sigma-C\Bigg),
\end{equation}
where, $L_{\mu}$ is the vessel length, $\beta$ is a Lagrange multiplier and $C$ is a constant \cite{Hu2013}. The first term corresponds to the total power dissipated in the network, and the second term imposes a metabolic or material cost characterized by $\sigma=1/\gamma-1$. For $\gamma=2/3$, this material cost is equal to the total volume of flow in the network, which is an important constraint for animal vasculature. 

We define the vessel mean-squared current at a given time $t'$ of adaptation as
\begin{equation}\label{eq3}
\langle Q_{\mu}^2 \rangle= \frac{1}{T}\int_0^T dt \hspace{5pt}\Big[\frac{1}{L_{\mu}}\int_0^{L_{\mu}} dz \hspace{5pt} Q_{\mu}(z,t)^2\Big],
\end{equation}
where $T=2\pi/\omega$ is the time-period of pulsatility. Note that we distinguish the adaptation time $t'$ in the AE (Eq.~\ref{eq1}) from the time used to calculate the vessel mean-squared current $t$, because adaptation typically occurs on much longer timescales than that of local modulations of flow in individual vessels (i.e. $t'\gg t$). 

When fluid inertia and vessel compliance are neglected, the current at node $N_1$ splits proportionally between the two vessels depending on their conductance, and the vessel mean-squared current has the form given in Fig.~\ref{fig1}(a). The critical transition of the AE for this two-vessel network $\forall \omega$ can be analytically shown to occur at $\gamma_c^{AE}=1/2$ (see supplemental materials). This is illustrated by the steady state flow-diagrams in $R_1-R_2$ space (Fig.~\ref{fig1}(b,c)). For $\gamma<\gamma_c^{AE}$, the diagonal has a stable fixed point (sink), which corresponds to a stabilized loop with vessels of equal radius at steady steady, irrespective of their initial sizes (Fig.~\ref{fig1}(b)). For $\gamma>\gamma_c^{AE}$, there exist two stable fixed points at the boundaries with large basins of attraction, indicating that for most initial conditions, one or the other vessel is lost. The diagonal has an unstable fixed point (saddle), indicating that the loop is stable only for a narrow range of initial conditions corresponding to exactly equal starting radii (Fig.~\ref{fig1}(b)). 

\section{Compliant Vessels}
The treatment above does not consider the opposition to changes in flow pressure due to fluid mass and the resulting fluid inertia. Moreover, biological networks are composed of compliant vessels, which can change in radius reversibly at short timescales to accommodate changes in flow volume. As shown in \cite{Barnard1966,Fancher2022}, inertia and compliance generates a finite time lag in flow propagation from the sources to the bulk of the transport network. Thus, in addition to the flow resistance (or conductance), the combined contribution of fluid inertia and compliance can be expected to alter the vessel mean-squared current on the timescale of adaptation. 
We follow the treatment of compliant vessels in \cite{Fancher2022}, and assuming an incompressible, laminar flow with rotational symmetry, the axial current $Q(z,t)$ and pressure $P(z,t)$ in each vessel satisfy
\begin{align}
\frac{\partial Q}{\partial z} &+c\frac{\partial P}{\partial t}=0,\label{eq4}\\
\frac{\partial P}{\partial z} + &l\frac{\partial Q}{\partial t} +rQ=0.\label{eq5}
\end{align}
The cross-section of the vessel changes as $A(z,t)=A_0 +cP(z,t)$ in response to wall pressure, where $c$ is the vessel compliance. We assume such changes in cross-section to be small in magnitude, i.e. $A_0 \gg cP(z,t)$. These vessel parameters can be combined to construct the characteristic length ($\lambda$), time ($\tau$) and admittance ($\alpha$) scales, which all vary proportional to the area of cross-section, as
\begin{align}
\lambda&=\lambda_0 (R/R_0)^2=\frac{2}{r}\sqrt{\frac{l}{c}},\nonumber\\
\tau&=\tau_0 (R/R_0)^2=\frac{2l}{r},\nonumber\\
\alpha&=\alpha_0 (R/R_0)^2=\sqrt{\frac{c}{l}},\label{eq6}
\end{align}
where $R_0$ is a typical radius. In particular, increasing $\lambda_0$  at constant radius, with $\alpha \lambda$ and $\tau$ held fixed, reflects a decrease in the compliance $c$ of the vessel. Generalizing the single compliant vessel to a network of compliant vessels is straightforward, and following \cite{Fancher2022}, the vessel mean-squared current $\langle Q_{\mu}^2 \rangle$ (Eq.~\ref{eq3}) can be calculated for each vessel. For the two-vessel network, this mean-squared current has a more complex dependence (see supplemental materials) on the vessel radius than the form given in Fig.~\ref{fig1}(a), and when used to drive the AE, results in a more realistic description of the evolution of the network structure, as we show below. For convenience, we will refer to the new framework of adaptation with fluid inertia and vessel compliance taken into consideration as the Modified Adaptation Equation (MAE), to distinguish it from the AE. 

\section{Results}
\begin{figure*}
\begin{center}
\includegraphics[scale=0.23]{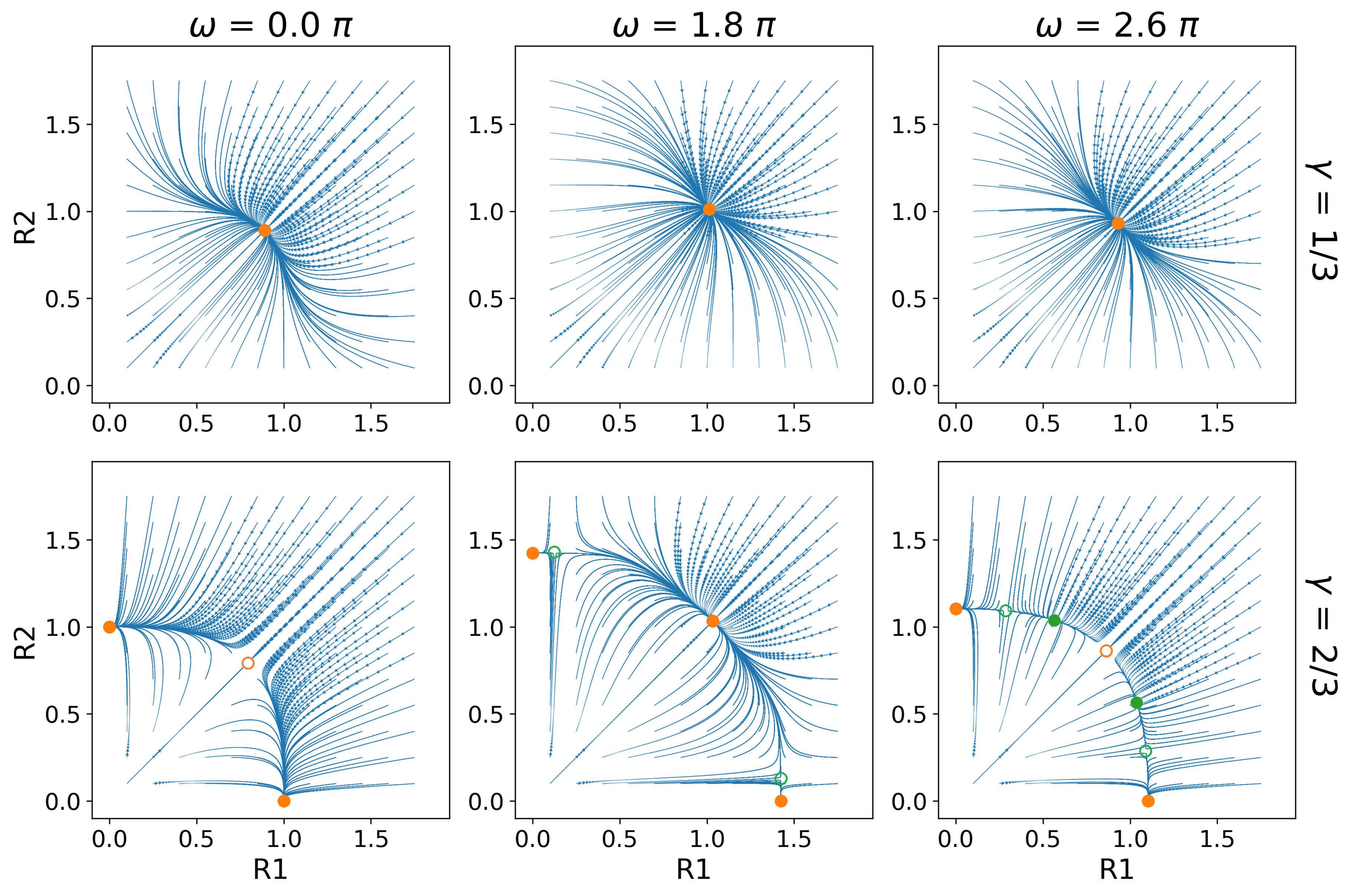}
\end{center}
\caption {\small{(color online) Flow Diagrams in $R_1-R_2$ space with $L_1=L_2=L$ and $L/\lambda_0<1$. Sinks denoted by solid circles and saddle points by hollow circles. For $\gamma=1/3$, the steady state is a stable loop with $R_1=R_2$ for all values of $\omega$. For ($\gamma=2/3,\omega=0$) loops are unstable for all initial conditions not on the diagonal. For ($\gamma=2/3,\omega=1.8\pi$) loops with vessels of equal radius are stable for almost all initial conditions not on the boundaries. For ($\gamma=2/3,\omega=2.6\pi$) loops with unequal vessel radii are stable for a broad range of intermediate initial conditions. Here $a=b=L=1$ and $\lambda_0=2$.}}
    \label{fig2}
\end{figure*}
\begin{figure}
\begin{center}
\includegraphics[scale=0.43]{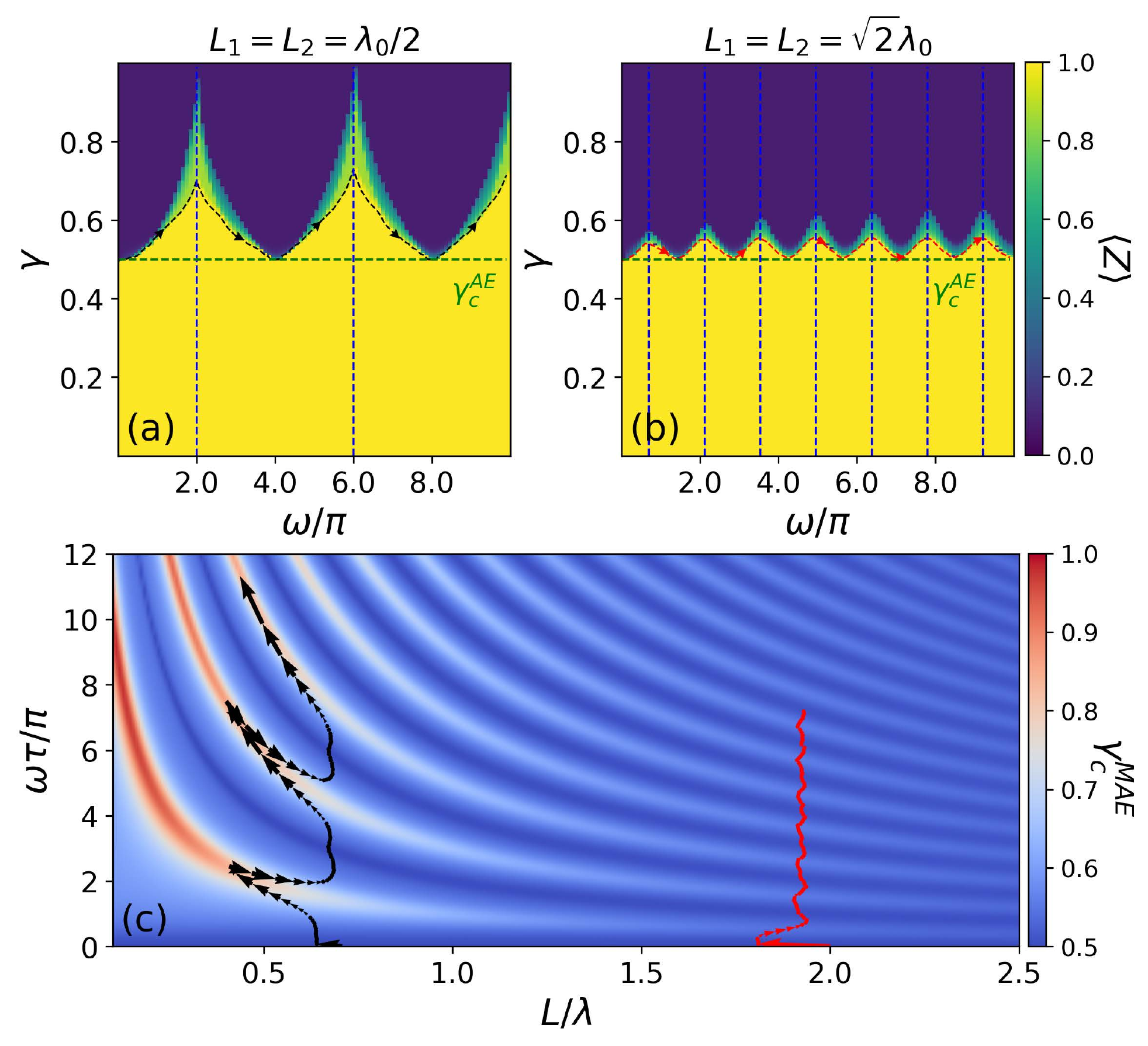}
\end{center}
\caption {\small{The critical transition for $L_1=L_2$. (a,b) Phase-diagram of $\langle Z\rangle$ in the $\gamma-\omega$ phase-space over the $144$ initial conditions in the range $R_1,R_2 \in (0.1,1.75)$ depicted in Fig.~\ref{fig2}. $L_1=L_2=1$ in (a), $L_1=L_2=2\sqrt{2}$ in (b), and $\lambda_0=2$ in both cases. Dashed green lines mark the critical transition in the AE, and dashed blue lines depict resonant frequencies. The black trajectory in (a) and the red trajectory in (b) follows $\gamma_c^{MAE}$ as a function of increasing $\omega$. (c) Phase-diagram of $\gamma_c^{MAE}$ in the $\omega\tau-L/\lambda$ phase-space. The black and red trajectories in (c) correspond to those in (a) and (b) respectively.}}
    \label{fig3}
\end{figure}
\begin{figure}
\begin{center}
\includegraphics[scale=0.43]{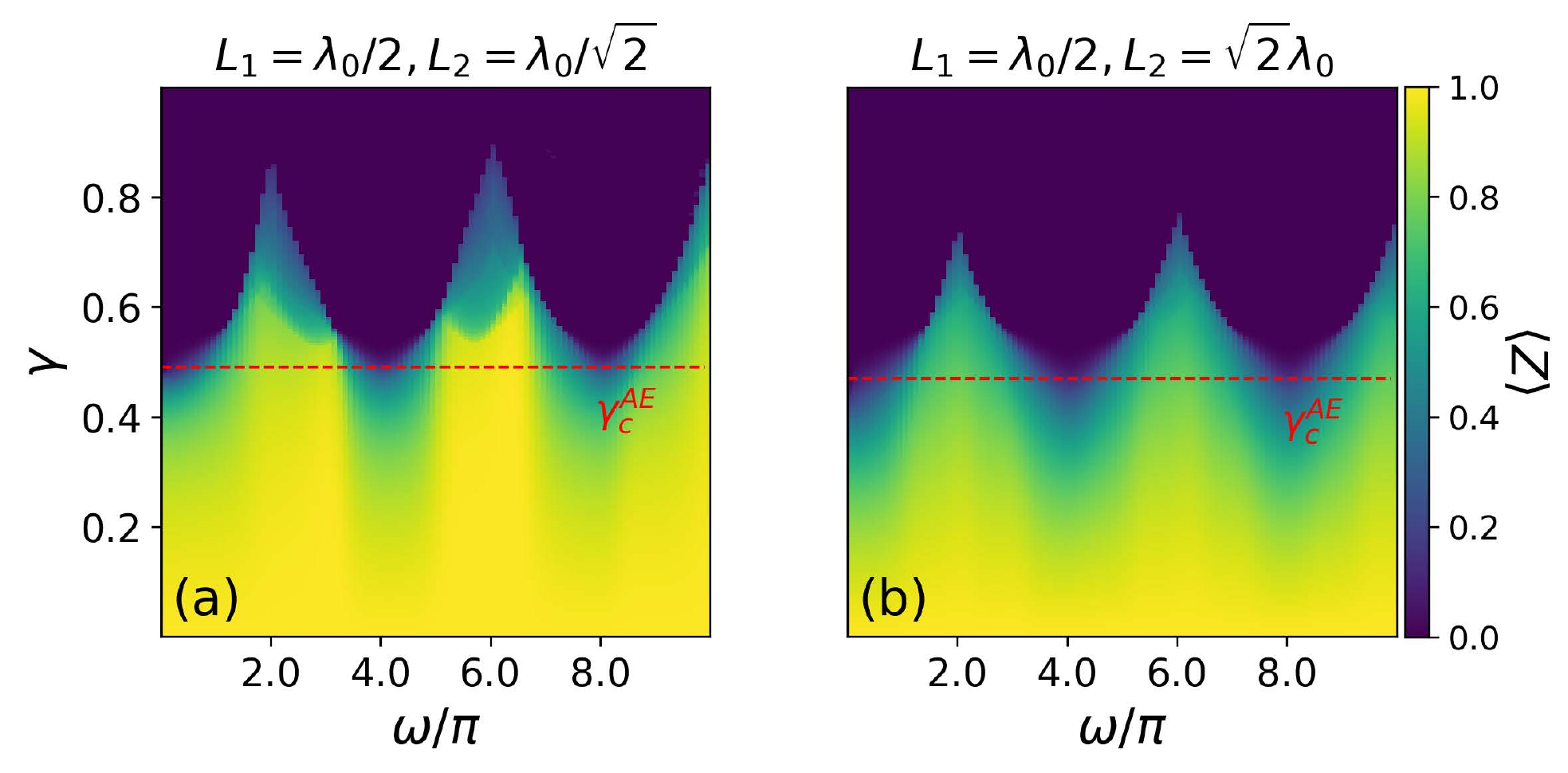}
\end{center}
\caption {\small{Phase-diagram of $\langle Z\rangle$ for $L_1 \neq L_2$. $L_1=1$ and $L_2=\sqrt{2}$ in (a), $L1=1$ and $L_2=2\sqrt{2}$ in (b), and $\lambda_0=2$ in both cases. Dashed red lines mark the critical transition in the AE.}}
    \label{fig4}
\end{figure}

The consequences of periodic driving of the MAE for the steady state structure of the two-vessel network are significant. As an example, Fig.~\ref{fig2} shows the steady state flow diagrams in $R_1-R_2$ space with vessels of equal but small effective lengths ($L/\lambda_0<1$), for which compliance is relatively low and pulsatile components of the flow are non-negligible. Like in the AE, the stable fixed point for $\gamma=1/3$ and all values of $\omega$, lies on the diagonal ($R_1=R_2$), indicating a stable symmetric loop. For $\gamma=2/3$, the non-pulsatile ($\omega=0$) flow diagram has sinks on the boundaries, meaning the loss of one or more vessel, once again similar to the AE result shown in Fig.~\ref{fig1}(c).
For $\omega=1.8\pi$ however, while stable steady states exist on both the diagonal as well as the boundaries, the former has a much broader basin of attraction than the latter. This implies that unlike the AE, for most initial conditions of the MAE loops can be stabilized for pulsatile driving at this frequency. Even more interestingly, for $\omega=2.6\pi$, new steady states with vessels of finite but unequal radii emerge and are stable, with basins of attraction larger than that of the boundary sinks. Clearly, at this frequency loops exist at steady state for a broad range of initial conditions, albeit with asymmetric flow distribution between the two vessels.

In general with pulsatility, the energy dissipation through vessel $\mu$ is amplified at special resonant frequencies depending on the value of $L_{\mu}/\lambda_0$, causing it to expand even for $\gamma>\gamma_c^{AE}$.
For each initial condition, we can calculate the quantity $Z=\min\Big[R_1/R_2,R_2/R_1\Big]_{t\to\infty}$, the minimum of the ratio of the radii of the vessels at steady state. For $Z=0$, the steady state is loopless, and for $0<Z \le 1$ the steady state is looped, with $Z=1$ corresponding to vessels of equal radius. The critical transition from a looped to a loop-less structure in the MAE as a function of the driving frequency can then be underpinned by the order parameter $\langle Z\rangle$, averaged over many initial conditions. Fig.~\ref{fig3}(a) shows the phase-diagram of $\langle Z\rangle$ in the $\gamma-\omega$ phase space, for the low compliance case depicted in Fig.~\ref{fig2}. The AE in this case would yield loops ($\langle Z\rangle>0$) below $\gamma_c^{AE}=1/2$ and no loops ($\langle Z\rangle=0$) above it (dashed red line). In contrast, the phase boundary between looped and loop-less steady states in the MAE has periodic modulations with respect to $\omega$, with loops stabilized for all physiologically relevant values of $\gamma$ at resonant frequencies (dashed blue lines). Moreover there is a significant spread of frequencies around the resonant values, for which loops are stable above $\gamma_c^{AE}$. Also, stable loops tend to be increasingly more asymmetric in radius for higher values of $\gamma$, even at resonant frequencies.

Increasing vessel lengths to be greater than $\lambda_0$, but still equal, decreases the looped $\langle Z \rangle>0$ phase in area, as shown in Fig.~\ref{fig3}(b). The periodic modulations with frequency in the critical transition are however retained, with shorter and more frequent peaks. Thus, even in the case of high relative compliance ($L/\lambda_0>1$), where the effect of pulsatile driving is significantly damped, the MAE is able to stabilize loops for a larger range of $\gamma$ values than the AE.

For $L_1=L_2=L$, an on-diagonal steady state always exists (Fig.~\ref{fig2}), but is stable for each $\omega$ value only below a critical value $\gamma=\gamma_c^{MAE}(\omega)$. Above $\gamma_c^{MAE}(\omega)$ the MAE generates either a loop-less network or an asymmetric loop at steady state ($0<\langle Z\rangle<1$). Fig~\ref{fig3}(c) shows the analytical phase diagram of $\gamma_c^{MAE}$ (see supplemental materials for derivation) in the $\omega\tau-L/\lambda$ space. Here $\tau$ and $\lambda$ are the full radius dependent time and length scales (Eq.~\ref{eq6}) corresponding to the on-diagonal fixed point for the given $\omega$ and for $\gamma=\gamma_c^{MAE}$. Clearly, the value of $\gamma_c^{MAE}$ oscillates with the frequency, with $\gamma_c^{AE}<\gamma_c^{MAE}\le 1$. The  trajectories in black and red in Fig~\ref{fig3}(c) correspond to the trajectories of the critical transition in Fig~\ref{fig3}(a,b), showing excellent agreement between the predictions of the numerical phase-diagram of $\langle Z\rangle=1$ and the analytical phase-diagram of $\gamma_c^{MAE}(\omega)$. The differences between the steady sate structures of the MAE and the AE are most pronounced for shorter vessels with $L/\lambda_0<1$, and the upper bound of oscillations in $\gamma_c^{MAE}(\omega)$ decreases monotonically with increasing $L/\lambda_0$.

Lastly, Fig.~\ref{fig4}(a,b) shows two cases with $L_1 \neq L_2$, one where both vessels are shorter, and another where only one vessel is shorter than $\lambda_0$. Here, unlike in the $L_1=L_2$ case, the on-diagonal steady state does not exist for all values of $\gamma$ and $\omega$. This explains the decrease in area of the $\langle Z\rangle=1$ phase from Fig.~\ref{fig3} to Fig.~\ref{fig4}. However, while the AE would predict $\langle Z\rangle=0$ for all $\gamma>\gamma_c^{AE}$ (dashed green line), clearly the MAE supports resonance frequencies (dashed blue lines) for which asymmetric loops are stabilized for a broad range of $\gamma$ values (see supplemental materials for a more detailed discussion).  

\section{Discussion} 
In summary, our simple model confirms that through the interplay between the time and length scales of pulsatile driving and those generated by network properties such as fluid inertia and vessel compliance, the modified adaptation framework (MAE) displays rich behavior that would be missed if the bulk flow was considered to instantaneously reflect modulations at the source, as in the AE. Loops (symmetric or asymmetric) are stable in the MAE for a much broader ranger of $\gamma$ values than in the AE, more so for less compliant vessels, that are shorter than the characteristic length scale $\lambda_0$. This is true even when the two vessels comprising the loop have unequal lengths, and importantly also when only one is shorter than the damping length scale.

It is crucial to understand the role of pulsatility in adaptation, because of its important consequences for the proper functioning and maintenance of biological transport networks. For instance, continuous-flow Left Ventricular Assist Devices (LVADs), used as a Bridge-To-Transplant therapy for advanced congestive heart failure, has been associated with detrimental pathology such as gastrointestinal bleeding, arterio-venous malformations and other complications, thought to stem from decreased arterial pulsatility \cite{Cheng2014,Bartoli2010}. Our finding that pulsatile driving can prevent the loss of vessels adaptation is at the very least consistent with such claims, even though our minimal model does not presume to capture the vast complexity of the human circulatory system.

Lastly, our results should be generalizable to mechanical networks \cite{hexnerPNAS2020,Bauerle2020} or bigger networks with many internal loops, for which the average energy dissipation over the network has been shown to be amplified at special resonant frequencies of the pulsatile driving \cite{Fancher2022}. The full analysis of the distribution of loops at steady state for such large, hierarchical networks is the scope of future work.

\begin{acknowledgments}
The authors acknowledge support from the NSF Award PHY-1554887, the Simons Foundation through Award 568888 and the University  of Pennsylvania  Materials  Research  Science and  Engineering  Center  (MRSEC)  through  Award  DMR-1720530.
\end{acknowledgments}

\bibliographystyle{apsrev4-1}
\bibliography{adaptation_compliance}

\pagebreak

\onecolumngrid
\appendix

\setcounter{equation}{0}
\setcounter{subsection}{0}
\setcounter{figure}{0}
\renewcommand{\theequation}{A\arabic{equation}}
\renewcommand{\thefigure}{A\arabic{figure}}
\renewcommand{\thesubsection}{A\arabic{subsection}}

\section*{Appendix}
In this appendix, we will derive the  expression for the vessel mean squared current $\langle Q_{\mu}^2 \rangle$ for the Modified Adaptation Equation (MAE), analyze the stability of the diagonal steady state in the case of vessels with equal lengths, and expand the discussion from the main text on the steady state structure for vessels with unequal lengths.
\subsection{Vessel mean squared current}\label{sec:Q2}
 Let $H_{\text{in}}(t)$ and $H_{\text{out}}(t)$ be the flow that goes into node $N_1$ and comes out from node $N_2$ respectively. Then $H_{in}=-H_{out}=Q_0+A\sin \omega t$, where $\omega=2\pi/T$ is the frequency of the pulsatile driving. This allows for three Fourier modes with $\tilde{H}^{0}=Q_0$, $\tilde{H}^{1}=\tilde{H}^{-1}=A/2$. For each vessel $e$, the solutions of Eq. $4$ and Eq. $5$ can thus be written as the discrete sum 

\begin{align}
Q_{\mu}(z,t) &= \sum_{n=-1}^{1}e^{in\omega_0t}\tilde{Q}^{(n)}_{\mu}(z),
\label{QFTF_disc}\\
P_\mu(z,t) &= \sum_{n=-1}^{1}e^{in\omega_0t}\tilde{P}^{(n)}_{\mu}(z),
\label{PFTF_disc}
\end{align}

where

\begin{align}
\tilde{Q}_{\mu}^{(n)}(z) &= \frac{in\omega\tau_{\mu}\alpha_{\mu}}{k(n\omega\tau_{\mu})}\frac{\tilde{P}^{(n)}_{in}\cosh\left(\frac{L_{\mu}-z}{\lambda_{\mu}}k(n\omega\tau_{\mu})\right)-\tilde{P}^{(n)}_{out}\cosh\left(\frac{z}{\lambda_{\mu}}k(n\omega\tau_{\mu})\right)}{\sinh\left(\frac{L_{\mu}}{\lambda_{\mu}}k(n\omega\tau_{\mu})\right)},
\label{QsolPbound_disc}\\
\tilde{P}_{\mu}^{(n)}(z) &= \frac{\tilde{P}^{(n)}_{in}\sinh\left(\frac{L_{\mu}-z}{\lambda_{\mu}}k(n\omega\tau_{\mu})\right)+\tilde{P}^{(n)}_{out}\sinh\left(\frac{z}{\lambda_{\mu}}k(n\omega\tau_{\mu})\right)}{\sinh\left(\frac{L_{\mu}}{\lambda_{\mu}}k(n\omega\tau_{\mu})\right)}.
\label{PsolPbound_disc}
\end{align}

Here $\tilde{P}^{\left(n\right)}_{in}=\tilde{P}_{\mu}^{\left(n\right)}(0)$ and $\tilde{P}^{\left(n\right)}_{out}=\tilde{P}_{\mu}^{\left(n\right)}(L_{\mu})$ are the boundary pressures, identical for each of the two vessels irrespective of their lengths. Eqs. $4$ and $5$ can be rearranged to take the form

\begin{equation}
\frac{\partial}{\partial z}\left(PQ\right)+\frac{1}{2}\frac{\partial}{\partial t}\left(\ell Q^{2}+cP^{2}\right)+rQ^{2} = 0.
\label{poweq}
\end{equation}

The mean squared current in each vessel is given by

\begin{align}
\langle Q_{\mu}^2 \rangle &=\frac{1}{rL_{\mu}T} \int_{0}^{T}dt\>\int_{0}^{L_{\mu}}dz\> Q_{\mu}(z,t)^{2} \nonumber\\
&= \frac{1}{rL_{\mu}T} \int_{0}^{T}dt\>\int_{0}^{L_{\mu}}dz\>\left(\frac{\partial}{\partial z}\left(P_{\mu}(z,t)Q_{\mu}(z,t)\right) + \frac{1}{2}\frac{\partial}{\partial t}\left(l\left(Q_{\mu}(z,t)\right)^{2} +c \left(P_{\mu}(z,t)\right)^{2}\right)\right), \nonumber\\
&=\frac{1}{rL_{\mu}T} \int_{0}^{T}dt\>\left( P_{\mu}(0,t)Q_{\mu}(0,t)-P_{\mu}(L_{\mu},t)Q_{\mu}(L_{\mu},t)\right) \nonumber\\
&\hspace{10pt}- \frac{1}{rL_{\mu}T} \int_{0}^{T}dt\> \left(\frac{\partial}{\partial t}\int_{0}^{L_{\mu}}dz\>\frac{1}{2}\frac{\partial}{\partial t}\left(l\left(Q_{\mu}(z,t)\right)^{2} +c \left(P_{\mu}(z,t)\right)^{2}\right)\right).
\label{dissint}
\end{align}

Noting that all quantities are identical at the start and end of each period, the second integral over time in Eq.~\ref{dissint} completely vanishes. We can then write the vessel mean squared current as a sum of Fourier modes as

\begin{align}\label{dissTave}
\langle Q_{\mu}^2 \rangle &= \frac{1}{rL_{\mu}T}\int_{0}^{T}dt\>\left(\left(\sum_{n'=-1}^{1}e^{in'\omega t}\tilde{P}^{(n')}_{\mu}(0)\right)\left(\sum_{n=-1}^{1}e^{in\omega t}\tilde{Q}^{(n)}_{\mu}(0)\right)\right.\nonumber\\ 
&\hspace{65pt} \left.- \left(\sum_{n'=-1}^{1}e^{in'\omega t}\tilde{P}^{(n')}_{\mu}(L_{\mu})\right)\left(\sum_{n=-1}^{1}e^{in\omega t}\tilde{Q}^{(n)}_{\mu}(L_{\mu})\right)\right), \nonumber\\
&= \frac{1}{rL_{\mu}}\sum_{n=-1}^{1}\left(\tilde{P}^{(-n)}_{\mu}(0)\tilde{Q}^{(n)}_{\mu}(0)-\tilde{P}^{(-n)}_{\mu}(L_{\mu})\tilde{Q}^{(n)}_{\mu}(L_{\mu})\right).
\end{align}

Eq. \ref{dissTave} can be expanded using Eq. \ref{QsolPbound_disc} and noting that $\tilde{P}^{(-n)}(z)=(\tilde{P}^{(n)}(z))^{*}$. With this, we get the expression for the vessel mean squared current in terms of the boundary pressures as 

\begin{align}
\langle Q_{\mu}^2 \rangle &= \frac{1}{rL_{\mu}} \sum_{n=-1}^{1}\frac{in\omega\tau_{\mu}\alpha_{\mu}}{k(n\omega \tau_{\mu})}\frac{\left(\abs{\tilde{P}^{(n)}_{in}}^{2}+\abs{\tilde{P}^{(n)}_{out}}^{2}\right)\cosh\left(\frac{L_{\mu}}{\lambda_{\mu}}k(n\omega\tau_{\mu})\right)-\tilde{P}^{(n)}_{in}\left(\tilde{P}^{(n)}_{out}\right)^{*}-\tilde{P}^{(n)}_{out}\left(\tilde{P}^{(n)}_{in}\right)^{*}}{\sinh\left(\frac{L_{\mu}}{\lambda_{\mu}}k(n\omega\tau_{\mu})\right)}, \nonumber\\
&= \left(\frac{\alpha_{\mu}\lambda_{\mu}}{2L_{\mu}}\right)^2\left(\tilde{P}^{(0)}_{in}-\tilde{P}^{(0)}_{out}\right)^{2} + \frac{\alpha_{\mu}\lambda_{\mu}}{L_{\mu}}\left(\abs{\tilde{P}^{(1)}_{in}}^{2}+\abs{\tilde{P}^{(1)}_{out}}^{2}\right)\text{Re}\left(\frac{i\omega\tau_{\mu}\alpha_{\mu}}{k(\omega\tau_{\mu})}\coth\left(\frac{L_{\mu}}{\lambda_{\mu}}k(\omega\tau_{\mu})\right)\right) \nonumber\\
&\quad -2\frac{\alpha_{\mu}\lambda_{\mu}}{L_{\mu}}\text{Re}\left(\tilde{P}^{(1)}_{in}\left(\tilde{P}^{(1)}_{out}\right)^{*}\right)\text{Re}\left(\frac{i\omega\tau_{\mu}\alpha_{\mu}}{k(\omega\tau_{\mu})}\csch\left(\frac{L_{\mu}}{\lambda_{\mu}}k(\omega\tau_{\mu})\right)\right).\label{dissexpand}
\end{align}

Thus, if we can calculate the boundary pressures $\tilde{P}^{(n)}_{in}$ and $\tilde{P}^{(n)}_{out}$ for the modes $n=0,1$, we can determine the mean squared current in each vessel at every step on the adaptation timescale. Denoting by These boundary pressures can be calculated by noting that

\begin{equation}\label{HPrel_TVN}
\begin{bmatrix}
\tilde{H}^{(n)}_{in} \\
\tilde{H}^{(n)}_{out}
\end{bmatrix} 
=
\begin{bmatrix}
\sum_{\mu} Q^{(n)}_{\mu}(0) \\
-\sum_{\mu} Q^{(n)}_{\mu}(L_{\mu})
\end{bmatrix} 
=\mathcal{L_{\mu}}(\omega)
\begin{bmatrix}
\tilde{P}^{(n)}_{in}(\omega) \\
\tilde{P}^{(n)}_{out}(\omega) 
\end{bmatrix},
\end{equation}

where $\mathcal{L_{\mu}}(\omega)$ is the network Laplacian, that takes the form

\begin{equation}\label{Lap_TVN}
\mathcal{L_{\mu}}(\omega) =
\begin{bmatrix}
\sum_{\mu}\frac{i\omega\tau_{\mu}\alpha_{\mu}}{k(\omega\tau_{\mu})}\coth\left(\frac{L_{\mu}}{\lambda_{\mu}}k(\omega\tau_{\mu})\right)
&-\sum_{\mu}\frac{i\omega\tau_{\mu}\alpha_{\mu}}{k(\omega\tau_{\mu})}\csch\left(\frac{L_{\mu}}{\lambda_{\mu}}k(\omega\tau_{\mu})\right)\\
-\sum_{\mu}\frac{i\omega\tau_{\mu}\alpha_{\mu}}{k(\omega\tau_{\mu})}\csch\left(\frac{L_{\mu}}{\lambda_{\mu}}k(\omega\tau_{\mu})\right)
&\sum_{\mu}\frac{i\omega\tau_{\mu}\alpha_{\mu}}{k(\omega\tau_{\mu})}\coth\left(\frac{L_{\mu}}{\lambda_{\mu}}k(\omega\tau_{\mu})\right)
\end{bmatrix}.
\end{equation}

In the $\omega=0$ case, the flow is steady and uniform, obeying

\begin{equation}
Q^{(0)}_{\mu} = \frac{\alpha_{\mu}\lambda_{\mu}}{2L_{\mu}}\left(\tilde{P}^{(0)}_{in}-\tilde{P}^{(0)}_{out}\right) = \frac{\alpha_{\mu}\lambda_{\mu}}{2L_{\mu}}\left(\frac{\alpha_{1}\lambda_{1}}{2L_1}+\frac{\alpha_{2}\lambda_{2}}{2L_2}\right)^{-1}\tilde{H}^{(0)} = \frac{R_{\mu}^{4}/L_{\mu}}{R_{1}^{4}/L_1+R_{2}^{4}/L_2}Q_0.
\label{Q0_TVN}
\end{equation}

In the pulsatile case, the imposed symmetry in the boundary currents along with the explicit form of the Laplacian also forces the relation $\tilde{P}^{(n)}_{in}(\omega)=-\tilde{P}^{(n)}_{out}(\omega)$. This readily admits the solution

\begin{align}\label{Psol_TVN}
\tilde{P}^{(n)}_{in}(\omega)=-\tilde{P}^{(n)}_{out}(\omega) &= \left(\sum_{\mu}\frac{i n\omega\tau_{\mu}\alpha_{\mu}}{k(n\omega\tau_{\mu})}\left(\coth\left(\frac{L_{\mu}}{\lambda_{\mu}}k(n\omega\tau_{\mu})\right)+\csch\left(\frac{L_{\mu}}{\lambda_{\mu}}k(n\omega\tau_{\mu})\right)\right)\right)^{-1}\tilde{H}^{(n)},\nonumber\\
&=\left(\sum_{\mu}\frac{i n\omega\tau_{\mu}\alpha_{\mu}}{k(n\omega\tau_{\mu})}\coth\left(\frac{L_{\mu}}{2\lambda_{\mu}}k(n\omega\tau_{\mu})\right)\right)^{-1}\tilde{H}^{(n)},
\end{align}

\noindent where we have used the relation $(\coth(x)+\csch(x))=\coth(x/2)$. Substituting Eq.~\ref{Q0_TVN} and Eq.~\ref{Psol_TVN} into Eq.~\ref{dissexpand}, we obtain the vessel mean squared current

\begin{align}
\langle Q_{\mu}^2\rangle &=Q_0^2\left(\frac{\alpha_{\mu}\lambda_{\mu}}{2L_{\mu}}\right)^2\left(\frac{\alpha_{1}\lambda_{1}}{2L_1}+\frac{\alpha_{2}\lambda_{2}}{2L_2}\right)^{-2} + \frac{2\alpha_{\mu}\lambda_{\mu}}{L_{\mu}}\abs{\tilde{P}^{(1)}_{in}(\omega)}^{2}\text{Re}\left(\frac{i\omega\tau_{\mu}\alpha_{\mu}}{k(\omega\tau_{\mu})}\left(\coth\left(\frac{L_{\mu}}{\lambda_{\mu}}k(\omega\tau_{\mu})\right)+\csch\left(\frac{L_{\mu}}{\lambda_{\mu}}k(\omega\tau_{\mu})\right)\right)\right) \nonumber\\
&=Q_0^2\left(\frac{\alpha_{\mu}\lambda_{\mu}}{2L_{\mu}}\right)^2\left(\frac{\alpha_{1}\lambda_{1}}{2L_1}+\frac{\alpha_{2}\lambda_{2}}{2L_2}\right)^{-2}+ \frac{A^2}{2}\frac{\frac{\alpha_{\mu}\lambda_{\mu}}{L_{\mu}}\text{Re}\left(\frac{i\omega\tau_{\mu}\alpha_{\mu}}{k(\omega\tau_{\mu})}\coth\left(\frac{L_{\mu}}{2\lambda_{\mu}}k(\omega\tau_{\mu})\right)\right)}{\abs{\sum_{\mu}\frac{i\omega\tau_{\mu}\alpha_{\mu}}{k(n\omega\tau_{\mu})}\coth\left(\frac{L_{\mu}}{2\lambda_{\mu}}k(n\omega\tau_{\mu})\right)}^2},
\label{aveQ_TVN}
\end{align}

Eq.~\ref{aveQ_TVN} can be simplified in notation by considering it as a function of $R_{1}$ and $R_{2}$. We first note that when Eq. $6$ of the main text is used to rewrite Eq.~\ref{aveQ_TVN}, the factors of $\alpha_{0}$ cancel out. We can then define the unitless function, $g(\omega,R_{\mu})$, and its zero frequency limit,

\begin{equation}
g(\omega,R_{\mu}) = \frac{i\omega\tau_{0}L_{\mu}}{k(\omega\tau_{0}(R_{\mu}/R_0)^2)\lambda_0}\coth\left(\frac{L_{\mu} R_0^2}{2\lambda_0R_e^2}k(\omega\tau_{0}(R_{\mu}/R_0)^2)\right) \quad\implies\quad \lim_{\omega\to 0}g(\omega,R_{\mu}) = 1.
\label{gdef}
\end{equation}

\noindent With this notation, Eq.~\ref{aveQ_TVN} can be written as

\begin{equation}\label{aveQ_gsimp}
\langle Q_{\mu}^2\rangle =
Q_0^2\left(\frac{R_{\mu}^{4}/L_{\mu}}{R_{1}^{4}/L_1+R_{2}^{4}/L_2}\right)^2 +
\frac{A^2}{2}\frac{(R_{\mu}^8/L_{\mu}^2)\text{Re}\left(g(\omega,R_{\mu})\right)}{\abs{(R_1^4/L_1)g(\omega,R_1)+(R_2^4/L_2)g(\omega,R_2)}^2}.
\end{equation}

The MAE then takes the form

\begin{equation}
\frac{dR_{\mu}}{dt'} = \frac{1}{R_{\mu}^{3}}\left(Q_0^2\left(\frac{R_{\mu}^{4}/L_{\mu}}{R_{1}^{4}/L_1+R_{2}^{4}/L_2}\right)^2 +
\frac{A^2}{2}\frac{(R_{\mu}^8/L_{\mu}^2)\text{Re}\left(g(\omega,R_{\mu})\right)}{\abs{(R_1^4/L_1)g(\omega,R_1)+(R_2^4/L_2)g(\omega,R_2)}^2}\right)^{\gamma}-bR_{\mu}.
\label{adaptcomb}
\end{equation}
\subsection{Stability of the Diagonal Fixed Point}\label{sec:diag_fp}
In this section, we will derive the critical value $\gamma_c^{MAE}(\omega)$ above which the diagonal steady state in the case of vessels with equal length ($L_1=L_2=L$) becomes unstable. To analyze the stability of the diagonal fixed point, i.e. to determine whether it is a sink or a saddle, we need merely to look at the eigenvalues of the exterior derivative matrix $M_{\mu\nu}=\partial(dR_{\mu}/dt)/\partial R_{\nu}$, where $\mu,\nu=1,2$. To begin constructing such a matrix, we first note that the derivative of $k(y)=\sqrt{iy(2+iy)}$ with respect to its argument is

\begin{equation}
\frac{d}{dy}\left(k\left(y\right)\right) = i\frac{1+iy}{k\left(y\right)} \quad\implies\quad \frac{d}{dy}\left(\frac{1}{k\left(y\right)}\right) = -\frac{i\left(1+iy\right)}{\left(k\left(y\right)\right)^{3}} = -\frac{1+iy}{y\left(2+iy\right)k\left(y\right)}.
\label{kder}
\end{equation}

\noindent This allows for the derivative of $g(\omega,R_{\mu})$ with respect to each radius, which we denote as $j(\omega,R_{\mu})$, to be evaluated as

\begin{align}
&j\left(\omega,R_{\mu}\right) = \frac{\partial}{\partial R_{\mu}}\left(g\left(\omega,R_{\mu}\right)\right) \nonumber\\
&= -\frac{i\omega\tau_{0}L_{\mu}}{\lambda_{0}}\cdot\frac{1+i\omega\tau_{0}\left(\frac{R_{\mu}}{R_0}\right)^{2}}{\omega\tau_{0}\left(\frac{R_{\mu}}{R_0}\right)^{2}\left(2+i\omega\tau_{0}\left(\frac{R_{\mu}}{R_0}\right)^{2}\right)k\left(\omega\tau_{0}\left(\frac{R_{\mu}}{R_0}\right)^{2}\right)}\cdot\frac{2\omega\tau_{0}R_{\mu}}{R_0^{2}}\cdot\coth\left(\frac{La_{0}^{2}}{2\lambda_{0}R_{\mu}^{2}}k\left(\omega\tau_{0}\left(\frac{R_{\mu}}{R_0}\right)^{2}\right)\right) \nonumber\\
&\quad -\frac{i\omega\tau_{0}L_{\mu}}{k\left(\omega\tau_{0}\left(\frac{R_{\mu}}{R_0}\right)^{2}\right)\lambda_{0}}\left(\frac{La_{0}^{2}}{2\lambda_{0}R_{\mu}^{2}}\cdot i\frac{1+i\omega\tau_{0}\left(\frac{R_{\mu}}{R_0}\right)^{2}}{k\left(\omega\tau_{0}\left(\frac{R_{\mu}}{R_0}\right)^{2}\right)}\cdot\frac{2\omega\tau_{0}R_{\mu}}{R_0^{2}}-\frac{La_{0}^{2}}{\lambda_{0}R_{\mu}^{3}}k\left(\omega\tau_{0}\left(\frac{R_{\mu}}{R_0}\right)^{2}\right)\right)\csch^{2}\left(\frac{La_{0}^{2}}{2\lambda_{0}R_{\mu}^{2}}k\left(\omega\tau_{0}\left(\frac{R_{\mu}}{R_0}\right)^{2}\right)\right) \nonumber\\
&= -\frac{2}{R_{\mu}}\cdot\frac{1+i\omega\tau_{0}\left(\frac{R_{\mu}}{R_0}\right)^{2}}{2+i\omega\tau_{0}\left(\frac{R_{\mu}}{R_0}\right)^{2}}\cdot\frac{i\omega\tau_{0}L_{\mu}}{k\left(\omega\tau_{0}\left(\frac{R_{\mu}}{R_0}\right)^{2}\right)\lambda_{0}}\coth\left(\frac{La_{0}^{2}}{2\lambda_{0}R_{\mu}^{2}}k\left(\omega\tau_{0}\left(\frac{R_{\mu}}{R_0}\right)^{2}\right)\right) \nonumber\\
&\quad +\frac{1}{R_{\mu}}\left(\frac{i\omega\tau_{0}L_{\mu}}{k\left(\omega\tau_{0}\left(\frac{R_{\mu}}{R_0}\right)^{2}\right)\lambda_{0}}\right)^{2}\csch^{2}\left(\frac{La_{0}^{2}}{2\lambda_{0}R_{\mu}^{2}}k\left(\omega\tau_{0}\left(\frac{R_{\mu}}{R_0}\right)^{2}\right)\right) \nonumber\\
&= \frac{1}{R_{\mu}}\left(\left(g\left(\omega,R_{\mu}\right)\right)^{2}-2g\left(\omega,R_{\mu}\right)\frac{1+i\omega\tau_{0}\left(\frac{R_{\mu}}{R_0}\right)^{2}}{2+i\omega\tau_{0}\left(\frac{R_{\mu}}{R_0}\right)^{2}}-\left(\frac{La_{0}^{2}}{\lambda_{0}R_{\mu}^{2}}\right)^{2}\frac{i\omega\tau_{0}\left(\frac{R_{\mu}}{R_0}\right)^{2}}{2+i\omega\tau_{0}\left(\frac{R_{\mu}}{R_0}\right)^{2}}\right)
\label{gder}
\end{align}

\noindent In the $\omega\to 0$ limit, $j(\omega,R_{\mu})$ vanishes, as can be readily determined by differentiating the $\omega\to 0$ limit of $g(\omega,R_{\mu})$ given in Eq. \ref{gdef}. Given this notation, we can readily differentiate $\langle Q_{\mu}^{2}\rangle$ with respect to $R_{\mu}$ to produce

\begin{align}
\frac{\partial}{\partial R_{\mu}}\left(\left\langle Q_{\mu}^{2}\right\rangle\right) &= Q_0^2\frac{R_{\mu}^{8}}{\left(R_1^{4}+R_2^{4}\right)^{2}}\left(\frac{8}{R_{\mu}}-\frac{8R_{\mu}^{3}}{R_1^{4}+R_2^{4}}\right)+ \frac{A^2}{2}\left(\frac{8R_{\mu}^{7}\text{Re}\left(g\left(\omega,R_{\mu}\right)\right)+R_{\mu}^{8}\text{Re}\left(j\left(\omega,R_{\mu}\right)\right)}{\abs{R_1^{4}g\left(\omega,R_1\right)+R_2^{4}g\left(\omega,R_2\right)}^{2}}\right. \nonumber\\
&\left.-\frac{2R_{\mu}^{8}\text{Re}\left(g\left(\omega,R_{\mu}\right)\right)\text{Re}\left(\left(4R_{\mu}^{3}g\left(\omega,R_{\mu}\right)+R_{\mu}^{4}j\left(\omega,R_{\mu}\right)\right)^{*}\left(R_1^{4}g\left(\omega,R_1\right)+R_2^{4}g\left(\omega,R_2\right)\right)\right)}{\abs{R_1^{4}g\left(\omega,R_1\right)+R_2^{4}g\left(\omega,R_2\right)}^{4}}\right) \nonumber\\
&= Q_0^2\frac{R_{\mu}^{8}}{\left(R_1^{4}+R_2^{4}\right)^{2}}\left(\frac{8}{R_{\mu}}-\frac{8R_{\mu}^{3}}{R_1^{4}+R_2^{4}}\right)+\frac{A^2}{2}\frac{R_{\mu}^{8}\text{Re}\left(g\left(\omega,R_{\mu}\right)\right)}{\abs{R_1^{4}g\left(\omega,R_1\right)+R_2^{4}g\left(\omega,R_2\right)}^{2}}\left(\frac{8}{R_{\mu}}+\frac{\text{Re}\left(j\left(\omega,R_{\mu}\right)\right)}{\text{Re}\left(g\left(\omega,R_{\mu}\right)\right)}\right. \nonumber\\
&\quad \left.-\frac{2\text{Re}\left(\left(4R_{\mu}^{3}g\left(\omega,R_{\mu}\right)+R_{\mu}^{4}j\left(\omega,R_{\mu}\right)\right)^{*}\left(R_1^{4}g\left(\omega,R_1\right)+R_2^{4}g\left(\omega,R_2\right)\right)\right)}{\abs{R_1^{4}g\left(\omega,R_1\right)+R_2^{4}g\left(\omega,R_2\right)}^{2}}\right).
\label{dQmdam}
\end{align}

\noindent Conversely, the derivative of $\langle Q_{\mu}^{2}\rangle$ with respect to $R_{\nu}$, where $\nu\ne\mu$, takes the form

\begin{align}
\frac{\partial}{\partial R_{\nu}}\left(\left\langle Q_{\mu}^{2}\right\rangle\right) &= -Q_0^2\frac{R_{\mu}^{8}}{\left(R_1^{4}+R_2^{4}\right)^{2}}\frac{8R_{\nu}^{3}}{R_1^{4}+R_2^{4}}\nonumber\\
&-A^2\frac{R_{\mu}^{8}\text{Re}\left(g\left(\omega,R_{\mu}\right)\right)\text{Re}\left(\left(4R_{\nu}^{3}g\left(\omega,R_{\nu}\right)+R_{\nu}^{4}j\left(\omega,R_{\nu}\right)\right)^{*}\left(R_1^{4}g\left(\omega,R_1\right)+R_2^{4}g\left(\omega,R_2\right)\right)\right)}{\abs{R_1^{4}g\left(\omega,R_1\right)+R_2^{4}g\left(\omega,R_2\right)}^{4}} \nonumber\\
&= -Q_0^2\frac{R_{\mu}^{8}}{\left(R_1^{4}+R_2^{4}\right)^{2}}\frac{8R_{\nu}^{3}}{R_1^{4}+R_2^{4}} \nonumber\\
&\quad-A^2\frac{R_{\mu}^{8}\text{Re}\left(g\left(\omega,R_{\mu}\right)\right)}{\abs{R_1^{4}g\left(\omega,R_1\right)+R_2^{4}g\left(\omega,R_2\right)}^{2}}\frac{\text{Re}\left(\left(4R_{\nu}^{3}g\left(\omega,R_{\nu}\right)+R_{\nu}^{4}j\left(\omega,R_{\nu}\right)\right)^{*}\left(R_1^{4}g\left(\omega,R_1\right)+R_2^{4}g\left(\omega,R_2\right)\right)\right)}{\abs{R_1^{4}g\left(\omega,R_1\right)+R_2^{4}g\left(\omega,R_2\right)}^{2}}.
\label{dQmdan}
\end{align}

\noindent Eqs. \ref{dQmdam} and \ref{dQmdan} then allow us to calculate the on diagonal and off diagonal matrix components $M_{\mu\mu}$ and $M_{\mu\nu}$ 

\begin{subequations}
\begin{align}
M_{\mu\mu} &= \gamma\frac{\left\langle Q_{\mu}^{2}\right\rangle^{\gamma-1}}{R_{\mu}^{3}}\left( Q_0^2\frac{R_{\mu}^{8}}{\left(R_1^{4}+R_2^{4}\right)^{2}}\left(\frac{8}{R_{\mu}}-\frac{8R_{\mu}^{3}}{R_1^{4}+R_2^{4}}\right)+\frac{A^2}{2}\frac{R_{\mu}^{8}\text{Re}\left(g\left(\omega,R_{\mu}\right)\right)}{\abs{R_1^{4}g\left(\omega,R_1\right)+R_2^{4}g\left(\omega,R_2\right)}^{2}}\left(\frac{8}{R_{\mu}}+\frac{\text{Re}\left(j\left(\omega,R_{\mu}\right)\right)}{\text{Re}\left(g\left(\omega,R_{\mu}\right)\right)}\right.\right. \nonumber\\
&\quad \left.\left.-\frac{2\text{Re}\left(\left(4R_{\mu}^{3}g\left(\omega,R_{\mu}\right)+R_{\mu}^{4}j\left(\omega,R_{\mu}\right)\right)^{*}\left(R_1^{4}g\left(\omega,R_1\right)+R_2^{4}g\left(\omega,R_2\right)\right)\right)}{\abs{R_1^{4}g\left(\omega,R_1\right)+R_2^{4}g\left(\omega,R_2\right)}^{2}}\right)\right)-3\frac{\left\langle Q_{\mu}^{2}\right\rangle^{\gamma}}{R_{\mu}^{4}}-b,
\label{Mmm}
\end{align}
\begin{align}
M_{\mu\nu} &= \gamma\frac{\left\langle Q_{\mu}^{2}\right\rangle^{\gamma-1}}{R_{\mu}^{3}}\left(-Q_0^2\frac{R_{\mu}^{8}}{\left(R_1^{4}+R_2^{4}\right)^{2}}\frac{8R_{\nu}^{3}}{R_1^{4}+R_2^{4}}\right. \nonumber\\
&\quad \left.-A^2\frac{R_{\mu}^{8}\text{Re}\left(g\left(\omega,R_{\mu}\right)\right)}{\abs{R_1^{4}g\left(\omega,R_1\right)+R_2^{4}g\left(\omega,R_2\right)}^{2}}\frac{\text{Re}\left(\left(4R_{\nu}^{3}g\left(\omega,R_{\nu}\right)+R_{\nu}^{4}j\left(\omega,R_{\nu}\right)\right)^{*}\left(R_1^{4}g\left(\omega,R_1\right)+R_2^{4}g\left(\omega,R_2\right)\right)\right)}{\abs{R_1^{4}g\left(\omega,R_1\right)+R_2^{4}g\left(\omega,R_2\right)}^{2}}\right).
\label{Mmn}
\end{align}
\label{Mels}
\end{subequations}

We are particularly interested in the case where $R_{\mu}=R_{\nu}=R$. This causes each mean squared current to reduce to

\begin{equation}
\left\langle Q^{2}\right\rangle = Q_0^2\frac{R^{8}}{\left(2R^{4}\right)^{2}}+\frac{A^2}{2}\frac{R^{8}\text{Re}\left(g\left(\omega,R\right)\right)}{\abs{2R^{4}g\left(\omega,R\right)}^{2}} = \frac{1}{4}Q_0^2+\frac{1}{2}\frac{A^2}{2}\text{Re}\left(\frac{1}{g\left(\omega,R\right)}\right).
\label{aveQ_equala}
\end{equation}

\noindent Additionally, this imposed symmetry forces $M_{11}=M_{22}$ and $M_{12}=M_{21}$. This particular form of $M$ always has $\lbrack 1,1\rbrack$ and $\lbrack 1,-1\rbrack$ as eigenvectors with eigenvalues we will denote as $\epsilon_{\parallel}$ and $\epsilon_{\perp}$ respectively. It is the sign of $\epsilon_{\perp}$ at the diagonal fixed point that dictates whether such a point is a stable basin or saddle. To investigate this, let $R^{*}$ be defined such that Eq. \ref{adaptcomb} vanishes when $R_{\mu}=R_{\nu}=R^{*}$, thus allowing for the relation $\langle Q^{2}\rangle^{\gamma}/{R^{*}}^{4}=b$. At this point, $\epsilon_{\perp}$ takes the form

\begin{align}
\epsilon_{\perp} &= \left(\gamma\frac{\left\langle Q^{2}\right\rangle^{\gamma-1}}{R^{3}}\left(Q_0^2\frac{R^{8}}{\left(2R^{4}\right)^{2}}\left(\frac{8}{R}-\frac{8a^{3}}{2R^{4}}\right)+\frac{A^2}{2}\frac{R^{8}\text{Re}\left(g\left(\omega,R\right)\right)}{\abs{2R^{4}g\left(\omega,R\right)}^{2}}\left(\frac{8}{R}\right.\right.\right. \nonumber\\
&\quad\quad \left.\left.\left.\left.+\frac{\text{Re}\left(j\left(\omega,R\right)\right)}{\text{Re}\left(g\left(\omega,R\right)\right)}-\frac{2\text{Re}\left(\left(4R^{3}g\left(\omega,R\right)+R^{4}j\left(\omega,R\right)\right)^{*}\left(2R^{4}g\left(\omega,R\right)\right)\right)}{\abs{2R^{4}g\left(\omega,R\right)}^{2}}\right)\right)-3\frac{\left\langle Q^{2}\right\rangle^{\gamma}}{R^{4}}-b\right)\right|_{R=R^{*}} \nonumber\\
&\quad -\left(\gamma\frac{\left\langle Q^{2}\right\rangle^{\gamma-1}}{R^{3}}\left(-Q_0^2\frac{R^{8}}{\left(2R^{4}\right)^{2}}\frac{8R^{3}}{2R^{4}}\right.\right. \nonumber\\
&\quad\quad \left.\left.\left.-A^2\frac{R^{8}\text{Re}\left(g\left(\omega,R\right)\right)}{\abs{2R^{4}g\left(\omega,R\right)}^{2}}\frac{\text{Re}\left(\left(4R^{3}g\left(\omega,R\right)+R^{4}j\left(\omega,R\right)\right)^{*}\left(2R^{4}g\left(\omega,R\right)\right)\right)}{\abs{2R^{4}g\left(\omega,R\right)}^{2}}\right)\right)\right|_{R=R^{*}} \nonumber\\
&= b\left(\frac{8\gamma}{\left\langle Q^{2}\right\rangle}\left(\frac{1}{4}Q_0^2+\frac{A^2}{8}\text{Re}\left(\frac{1}{g\left(\omega,R^{*}\right)}\right)\left(1+\frac{\text{Re}\left(R^{*}j\left(\omega,R^{*}\right)\right)}{8\text{Re}\left(g\left(\omega,R^{*}\right)\right)}\right)\right)-4\right) \nonumber\\
&= 4b\left(2\gamma\frac{Q_0^2+\frac{A^2}{2}\text{Re}\left(\frac{1}{g\left(\omega,R^{*}\right)}\right)\left(1+\frac{\text{Re}\left(R^{*}j\left(\omega,R^{*}\right)\right)}{8\text{Re}\left(g\left(\omega,R^{*}\right)\right)}\right)}{Q_0^2+\frac{A^2}{2}\text{Re}\left(\frac{1}{g\left(\omega,R^{*}\right)}\right)}-1\right).
\label{eigen_perp}
\end{align}

\noindent From Eq. \ref{eigen_perp} it is clear that the value of $\gamma$ directly determines the sign of $\epsilon_{\perp}$. The transition value of $\gamma$ at which $\epsilon_{\perp}=0$ takes the form

\begin{equation}
\gamma_c^{MAE}(\omega) = \frac{1}{2}\frac{Q_0^2+\frac{A^2}{2}\text{Re}\left(\frac{1}{g\left(\omega,R^{*}\right)}\right)}{Q_0^2+\frac{A^2}{2}\text{Re}\left(\frac{1}{g\left(\omega,R^{*}\right)}\right)\left(1+\frac{\text{Re}\left(R^{*}j\left(\omega,R^{*}\right)\right)}{8\text{Re}\left(g\left(\omega,R^{*}\right)\right)}\right)}.
\label{gamma_perp}
\end{equation}

The case of the traditional adaptation equation (AE) occurs when $A=0$, which in turn causes Eq. \ref{gamma_perp} to simply become $\gamma_{c}^{AE}=1/2$. For the modified adaption equation considered in this work with nonzero $\omega$, the steady state is a symmetric loop with equal vessel radii for $\gamma<\gamma_c^{MAE}(\omega)$, and either loopless or an asymmetric loop with unequal vessel radii for $\gamma>\gamma_c^{MAE}(\omega)$. Fig.~$3$(c) of the main text can be constructed by plotting the value of $\gamma_c^{MAE}$ given in Eq.~\ref{gamma_perp} for each combination of $\omega\tau=\omega \tau_0 (R^{*}/R_0)^2$ and $L/\lambda=L/(\lambda_0(R^{*}/R_0)^2)$, where $R^{*}$ is the steady state radius of each vessel at that specific value of $\omega$ and at $\gamma=\gamma_c^{MAE}(\omega)$.
\subsection{Vessels of unequal lengths}\label{sec:unequal}
\begin{figure}
\begin{center}
\includegraphics[scale=0.75]{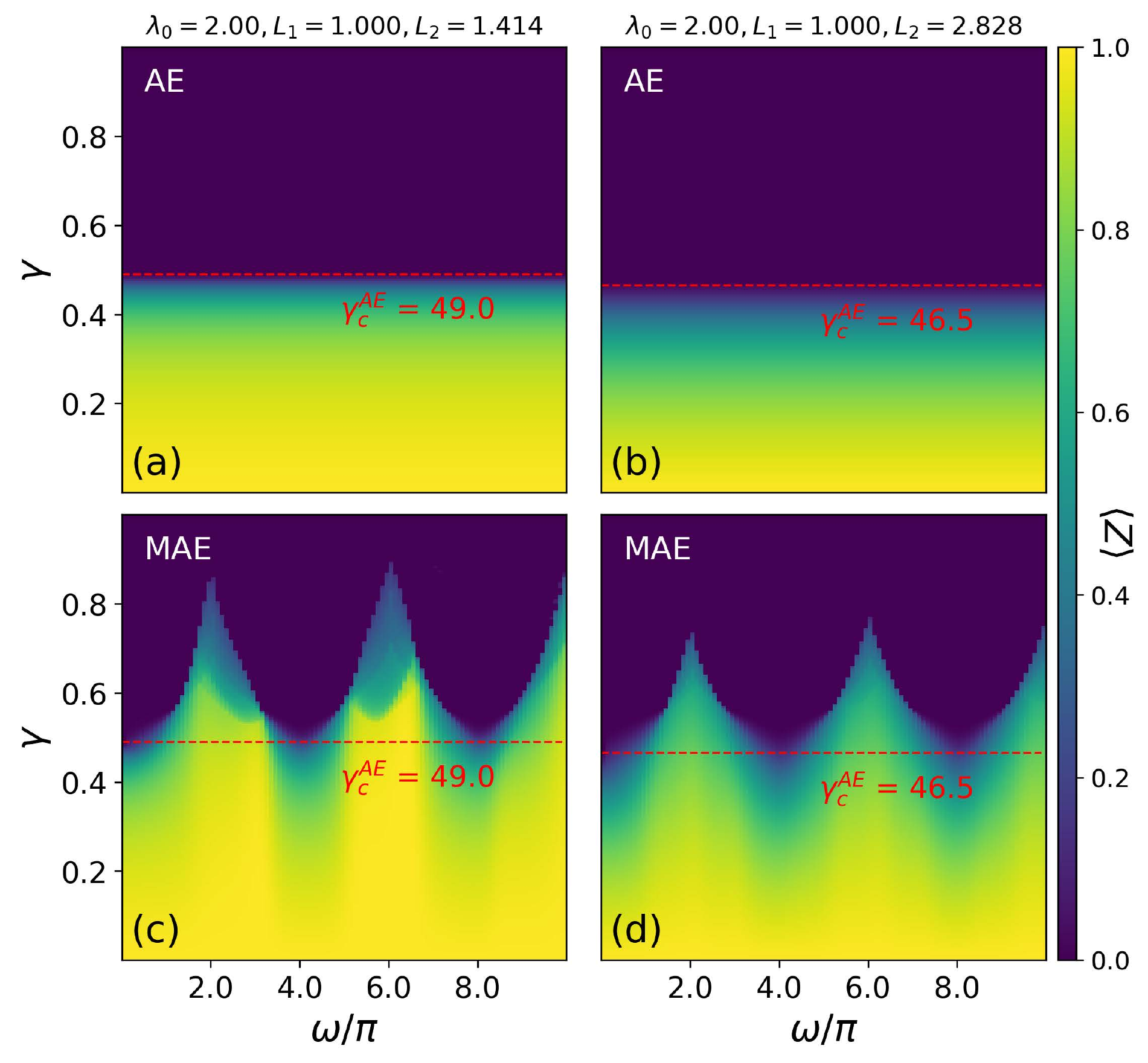}
\end{center}
\caption {\small{(color online) Comparison of the steady state structures in the AE and MAE for vessels of unequal lengths. Top: Phase-diagrams in the AE, with $L_1=1$ and $L_2=\sqrt{2}$ in (a), and $L_1=1$ and $L_2=2\sqrt{2}$ in (b). Bottom: Phase-diagrams in the MAE, with $L_1=1$ and $L_2=\sqrt{2}$ in (c), and $L_1=1$ and $L_2=2\sqrt{2}$ in (d). $\lambda_0=2$ in all four cases, and dashed red lines depict the critical transition in the AE for each case.}}
    \label{figS1}
\end{figure}

In this section we will discuss the steady state behavior obtained for cases where the two vessels have unequal lengths. Fig.~\ref{figS1} shows the phase-diagram of the order parameter $\langle Z\rangle$ obtained with the AE (top panels) and the MAE (bottom panels) for the two cases discussed in Fig.~$4$ of the main text. While in both cases, the second vessel is longer that the first vessel, in the first case both vessels are shorter than the characteristic length scale $\lambda_0$ (Fig.~\ref{figS1}(a,c)), whereas in the second case, the first vessel is shorter and the second vessel longer than $\lambda_0$ (Fig.~\ref{figS1}(b,d)). It is clear that in either scenario, the MAE stabilizes loops for a larger range of $\gamma$ values than the AE. We also observe that as the length of the second vessel is increased in comparison to the first, resonances become weaker, and the peaks in $\langle Z\rangle$ values become shorter in height. 

In contrast to the situation where vessels have equal lengths the transition from a looped to a loopless steady state is not sharp either the AE for unequal vessel lengths. Instead we observe a more gradual transition from $\langle Z\rangle=1$ to $\langle Z\rangle= 0$ in Fig.~\ref{figS1}(a,b). This is because unlike the special symmetry furnished by the equal vessel case, vessels of different lengths donot always support a steady state on the diagonal of the $R_1-R_2$ space, for which $\langle Z\rangle$ would be unity. Instead, asymmetric loops with $0<\langle Z\rangle <1$, i.e. with different but non-zero radii for each vessel, are stabilized below $\gamma_c^{AE}$ in the AE. This also explains why the MAE phase-diagram of $\langle Z\rangle$ (Fig.~\ref{figS1}(c,d)) has less sharp boundaries between the yellow ($\langle Z\rangle=1$) and dark blue regions ($\langle Z\rangle=0$) than in the equal vessel cases shown in Fig.~$3$ of the main text.

\begin{figure}
\begin{center}
\includegraphics[scale=0.65]{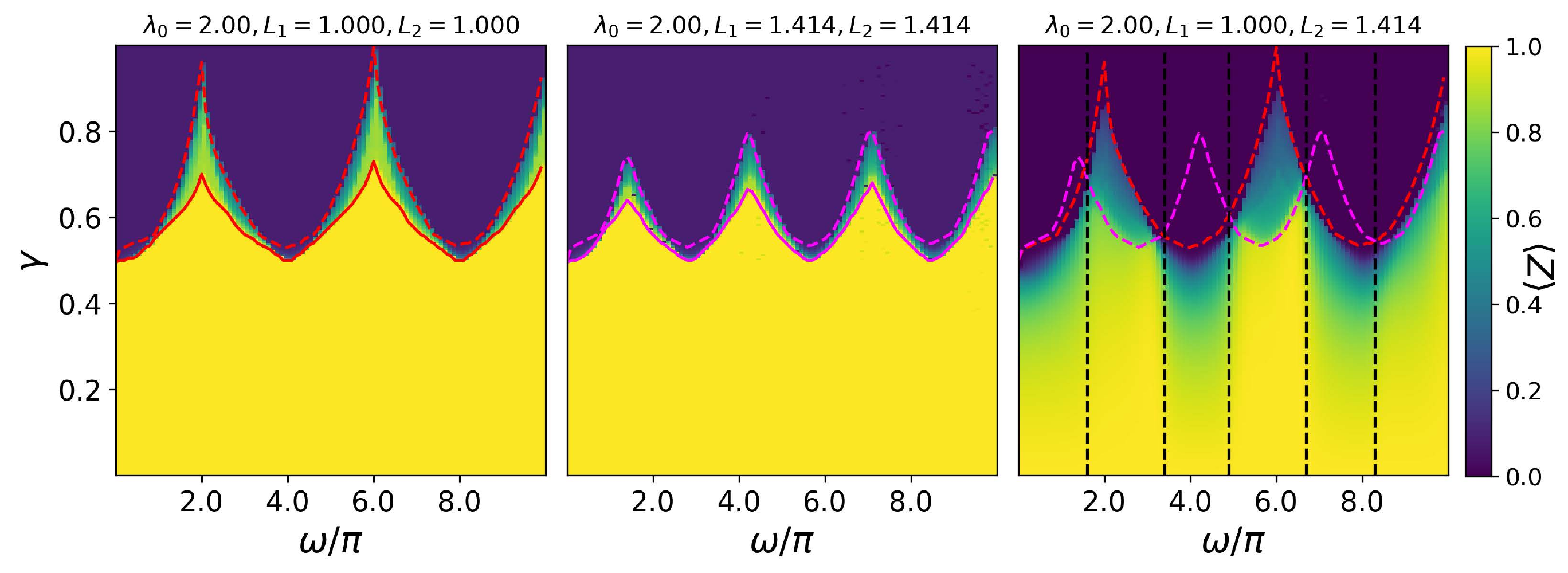}
\end{center}
\caption {\small{(color online) Steady state structure for unequal vessel lengths compared to the corresponding equal vessel length cases. (a,b) Phase-diagram for equal vessel lengths, with $L=1$ in (a) and $L=\sqrt{2}$ in (b), and $\lambda_0=2$. Solid lines depict the values of $\gamma(\omega)$ above which symmetric loops ($\langle Z\rangle=1$) become unstable. Dashed lines depict the values of $\gamma(\omega)$ above which asymmetric loops ($\langle Z\rangle>0$) become unstable. (c) Phase-diagram for the unequal vessel length case with $L_1=1$ and $L_2=\sqrt{2}$. The colored dashed lines correspond to those in (a,b). Vertical black dashed lines demarcate regions of the $\omega-\gamma$ phase space that have the maximum values of the order parameter $\langle Z\rangle$, that is the most symmetric loops at steady state.}}
    \label{figS2}
\end{figure}

The qualitative features of the phase-diagram of $\langle Z\rangle$ for unequal vessel lengths $L_1\ne L_2$ can be understood by looking at the superposition of the cases where both vessels have length $L=L_1$ or $L=L_2$. Fig.~\ref{figS2} shows such a comparison for vessels shorter than $\lambda_0$. In the equal vessel length cases (Fig.~\ref{figS2}(a,b)) we have marked with solid lines the values of $\gamma(\omega)$ above which symmetric loops ($\langle Z\rangle=1$) become unstable, and with dashed lines the values of $\gamma(\omega)$ above which even asymmetric loops become unstable. Firstly, it is clear from Fig.~\ref{figS2}(c), that the resonances of the shorter vessel dominates the phase diagram in the unequal vessel lengths case, generating peaks of $\langle Z\rangle$ at the same frequencies as in Fig.~\ref{figS2}(a). Secondly, the regions of the $\omega-\gamma$ phase space which supports the highest values of $\langle Z\rangle\approx 1$, is the region for which the steady state is looped, i.e $\langle Z\rangle>0$ for each of the equal vessel length cases (demarcated by dashed black lines). In regions where only one or the other equal vessel length cases is looped, a sort of destructive interference leads to a reduction in the value of the order parameter $\langle Z\rangle$ for the unequal vessel length case. Thus, even without calculating the full phase-diagram for a network with a mixture of short and long vessels, we should be able to qualitatively predict its steady state structure from the corresponding equal vessel length cases.

\end{document}